# Beam Geometry-Controlled Nonequilibrium Formation of WS$_2$/CsPbBr$_3$ Hybrids and Interfacial Carrier Dynamics

Challa Rajendra Kumar,[1] M.S.S. Bharathi,[1] Rahul Murali,[1]
Venugopal Rao Soma,[2,*] Sai Santosh Kumar Raavi[1,*]

[1]Ultrafast Photophysics and Photonics Laboratory, Department of Physics, Indian Institute of Technology Hyderabad, Kandi, 502285, Telangana, India

[2]School of Physics and DRDO Industry Academia – Centre of Excellence (formerly ACRHEM), University of Hyderabad, Hyderabad 500046, Telangana India

**Authors for Correspondence:** soma_venu@uohyd.ac.in and sskraavi@phy.iith.ac.in

**ABSTRACT:** Perovskite based nanocomposites provide a versatile platform for tailoring light matter coupling and carrier injection in low dimensional optoelectronics. In particular hybrid structures incorporating transition metal dichalcogenides (TMDCs) enable tunable band alignment and excitonic dynamics. However, the central challenge lies in the scalable synthesis of defect controlled layered TMDCs suitable for coherent interfacial coupling. Here we demonstrate geometry dependent femtosecond (*fs*) laser ablation in liquid as a scalable strategy for defect controlled exfoliation of WS$_2$ and in situ formation of WS$_2$/CsPbBr$_3$ nanocomposite. Using 50 *fs* pulses at identical fluence, we directly compare Gaussian and Bessel beam profiles to elucidate how spatial energy distribution governs nonlinear carrier generation, Coulomb stress and electron-lattice energy transfer. Multiphysics modeling combining nonlinear absorption, Coulomb instability criteria and two temperature dynamics reveals that the axially extended, weakly diffracting Bessel profile excites carrier densities above the electronic destabilization threshold while reducing the peak lattice heating. In contrast, Gaussian excitation concentrates energy within the focal volume, promoting rapid thermalization and defect formation. Spectroscopic analysis reveals reduced trap assisted recombination and enhanced excitonic stability in Bessel processed samples. Transient absorption spectroscopy (TAS) further demonstrates prolonged carrier lifetime. Extending this approach, single step ablation yields WS$_2$/CsPbBr$_3$ hybrids exhibiting Type-I band alignment and geometry dependent interfacial carrier injection, evidenced by photoluminescence quenching and modified ultrafast decay dynamics. These results establish spatial beam shaping as a scalable control parameter for ultrafast synthesis of defect free TMDCs and nanocomposites. Together these findings provide insight into geometry dependent ablation dynamics and establish process guidelines for next generation optoelectronic materials.

# I. INTRODUCTION

The rapid evolution of nanotechnology demands materials on atomic scale with tunable electronic, optical properties with mechanical stability (1). While silicon dominates in microelectronics and graphene offers exceptional mobility, their lack of a tunable bandgap limits their utility in advanced logic and optoelectronic devices (2). Hybrid nanocomposites (NCs) constructed from low-dimensional materials provides a route for emergent interfacial phenomena that are not accessible in individual components (3), solving the above limitations. By integrating with complementary electronic structures and optical response, NCs enable enhanced charge transfer, exciton dissociation, band alignment control and improved stability (4). Such interfacial design has become central to next generation optoelectronics, photonics applications (5). Among the hybrid systems perovskite based NCs have attracted significant attention due to exceptional optoelectronic properties including strong optical absorption, long carrier diffusion lengths, defect tolerance and high photoluminescence quantum yield (6). Despite these advantages, they suffer from environmental stability, ion migration, and interfacial recombination losses. Integrating perovskites with two-dimensional semiconductors provides an effective strategy to enhance carrier extraction, suppress nonradioactive recombination and stabilize exciton dynamics through interfacial coupling (7). Transition Metal Dichalcogenides (TMDC), atomically thick class of materials belonging to 2D family, are promising candidates for such hybridization. They offer intrinsic semiconducting properties, strong light-matter interactions, and compatibility with flexible, transparent platforms making them highly suitable for next-generation electronic and photonic applications (8). Among them layered TMDC's, possessing high surface to volume ratio, further enhance the surface interaction offering a unique combination of advantages that are not observed in any conventional 3D materials (9) (10). The creation of such high-quality NCs with TMDCs, is crucial for a wide range of advanced applications, including electronics, energy storage, and sensors.

Despite the promising properties of layered TMDCs, their wafer-scale synthesis with consistent quality remains a major challenge (8, 11). In bulk TMDCs, strong intralayer covalent bonding is coupled with weak interlayer van der Waals (vdW) interaction enabling exfoliation under relatively small external stimuli. However, the size confinement makes challenging for controlled exfoliation (12, 13). Achieving precise control to produce defect free material with consistent quality remains hurdle (14). Mechanical exfoliation, which is employed widely in research to obtain high quality monolayers, suffers from limitations, it is a low-throughput method, yields flakes of random size and position, and it is unsuitable for scalable manufacturing (15, 16). Chemical vapor deposition (CVD), the most common method for large-area growth, suffers from non-uniform thickness control, grain boundary formation, and defect densities that degrade charge transport and optical performance (17, 18). Liquid phase exfoliation techniques based on probe sonication or shear mixing produce high yield but relay on prolonged mechanical agitation (19, 20). Chemical and electrochemical intercalation methods weaken interlayer coupling but induces phase instability, chemical contamination, or irreversible lattice distortion (21, 22). Together these limitations highlight a fundamental gap between material quality and scalability that conventional exfoliation techniques struggle to bridge (23), particularly for hybrid TMDC- perovskite NC systems where interfacial cleanliness is important.

Femtosecond laser ablation in liquids has emerged as a promising contamination free technique, alternative capable of scalable material processing (24, 25). While laser ablation in liquid has been extensively studied for nanoparticle synthesis, its application to TMDC exfoliation remains comparatively less explored. This method fundamentally relies on depositing optical energy into the material on a time scales of ($10^{-15}$ s) much shorter than heat diffusion or lattice equilibrium (26). During this ultrashort interaction, energy is absorbed by electrons leading to the creation of highly non equilibrium electronic state before the lattice responds thermally (27). This transient electronic excitation weakens the interatomic bonding and removes the material before significant heat accumulation, thereby limiting the thermal damage. For layered materials such as TMDCs, this ultrafast and localized electronic excitation provides an effective pathway for exfoliation of weakly Vander Waal bound adjacent layers. Rather than relying on slow thermal expansion or chemical intercalation, the process transiently destabilizes interlayer bonding, enabling layer separation while preserving in plane crystallinity. However, the challenges in laser-based material processing, are to minimize the heat affected zone, while achieving high surface quality without compromising exfoliation efficiency. Ultrafast ablation is governed by many parameters, like laser pulse duration, wavelength, fluence and repetition rate which collectively determines energy coupling, thermal diffusion or material removal dynamics (28). Extensive efforts have given on optimizing these parameters to suppress thermal damage and improve precession(29, 30). Beyond these intrinsic pulse characteristics, the other degree of freedom lies in structuring the spatial profile of beam itself.

In practice most laser-based exfoliation studies employ fundamental Gaussian beam (GB) profiles for ablation, drilling and exfoliation etc., where optical energy is strongly localized both laterally and axially, as visualized in fig. 1(a) (31). Despite its wide spread use, it intrinsically suffers from strong spatial energy localization, a limited focal volume defined by the Rayleigh range, and pronounced nonlinear distortion at high fluence. These effects lead to non-uniform energy deposition, steep thermal gradients and low aspect ratio interaction region, constraining control over ablation efficiency and material quality. Recent studies have therefore turned toward a new structured light, Bessel beam (BB), which are the solution invariant of the paraxial wave equation (32, 33). BB are characterized by a narrow central core surrounded by concentric side lobes that replenish continuously on axis as in fig. 1(b), unlike GB. The spatial distribution of optical intensity and its propagation behavior which governs the energy deposition in the material distinct the BB over GB. A fundamental GB exhibit transverse intensity profile given by(34),

$$I_G(r,z) = I_0 \left(\frac{w_0}{w(z)}\right)^2 \exp\left(-\frac{2r^2}{w^2(z)}\right), \tag{1}$$

Where $w(z) = w_0\sqrt{1+(z/z_R)^2}$ is the beam radius and $z_R$ is the Rayleigh range. This profile implies strong axial diffraction and divergence away from the focal plane, resulting in localized deposition confined to smaller cross-section. When the fundamental GB of eqn. 1 is passed through the axicon lens it produces a BB profile of intensity distribution,

$$I_B(r) \propto J_0^2(k_r r), \tag{2}$$

an ideal zeroth-order BB is described by a transverse electric field

$$E_B(r,z) = E_0 J_0(k_r r) e^{ik_z z}, \tag{3}$$

Where $J_0$ is the zeroth order Bessel function, and $k_r$ and $k_z$ are the transverse and longitudinal wave vectors, respectively. These function exhibit propagation invariant intensity profiles over extended axial distance. The surrounding concentric side lobes acts as an effective energy reservoir that continuously replenishing the central core, giving rise to the self-healing property, producing a long, quasi uniform interaction volume. We believe that this intrinsic difference in spatial energy distribution has profound implications in ultrafast laser matter interaction.

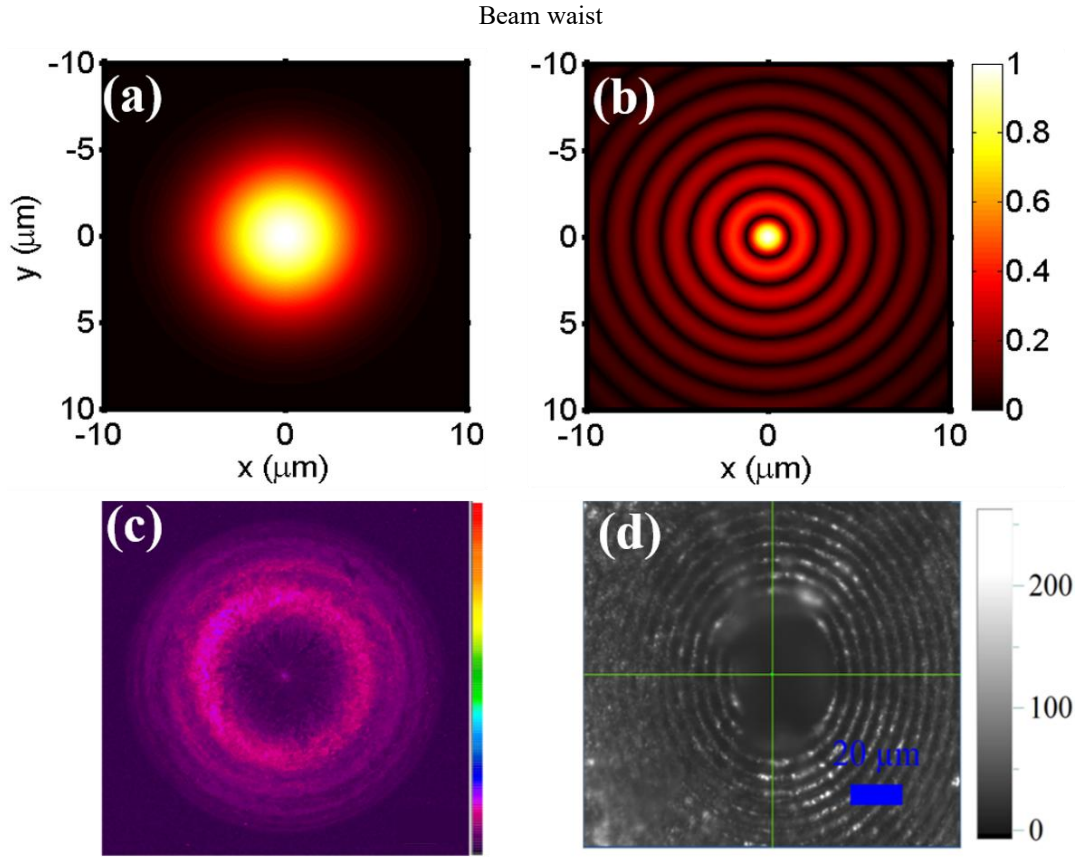

**FIG. 1.** : (a), (b) Simulated intensity profile for Gaussian, strongly confined near the axis and Bessel beam surrounded by concentric side lobes described by $J_0^2(k_r r)$ (c) Beam profile of a Bessel beam as measured by profilometer (d) Formation of concentric rings on a $WS_2$ pellet after irradiation with the Bessel beam at a fixed location for few seconds, clearly demonstrating the characteristic pattern of the Bessel beam.

Furthermore, in liquid environments, refractive index fluctuations and scattering from cavitation bubbles often destabilize Gaussian focal spot, leading to variable energy delivery (35). The self-construction nature of BB mitigates this instability, maintaining a stable excitation geometry even in highly scattering media. The advantage of BB is not only their extended depth of focus, but their ability to decouple electronic excitation from lattice heating through the spatial delocalized energy deposition.

Herein we report femtosecond laser ablation-in-liquid strategy for exfoliation of few-layer $WS_2$ and simultaneously producing perovskite NC, enabling scalable, high quality 2D-$WS_2$ nano sheets and WS$_2$/CsPBbr$_3$ synthesis. We systematically investigate beam profile dependent ultrafast dynamics by comparing Gaussian and Bessel excitation under identical fluence conditions. To understand the origin of the excitation geometry dependence, we employed the nonlinear carrier generation calculation, Coulomb stress analysis and two temperature modeling to describe electron-lattice non equilibrium and competing thermal and non-thermal ablation pathways. By tailoring the spatial intensity distribution, we demonstrate that the beam geometry governs the carrier excitation, electronic pressure and subsequent lattice response, thereby dictating exfoliation efficiency and defect formation. Optical and morphological studies are explored to investigate the structural and luminescent properties of ablated WS$_2$. Furthermore, perovskite TMDC nanocomposite is engineered to collate interfacial charge transfer, leveraging favorable band alignment for carrier injection. Transient absorption spectroscopy (TAS) reveals distinct carrier relaxation pathways, enabling comparison of exciton dynamics, trapping and recombination processes in pristine and hybrid systems. This beam geometry driven control over electron lattice disequilibrium introduces a new physical handle for steering ultrafast ablation mechanisms, enabling efficient exfoliation and NC formation, while suppressing the formation of interfacial defects.

## II. Beam profile dependent ablation dynamics

Under *fs* excitation of WS$_2$, the optical energy is deposited into the electronic subsystem on a time scale of *50 fs*, where the system is driven far from thermodynamic equilibrium immediately after irradiation (30). In this ultrafast regime, material ablation is governed by two dominant mechanisms: electronically driven (Columbic) non thermal ablation (36), in which rapid carrier generation creates a transient space charge field and associated electronic pressure capable of destabilizing interlayer bonding before significant lattice heating, and the second is thermally driven lattice destabilization (37), where absorbed energy is transferred to the lattice via electron–phonon (e-ph) coupling, resulting in superheating and explosive vaporization of material. In ablation, both mechanisms are activated however to separate Coulomb driven mechanism and confirm the material removal, the following conditions must be simultaneously satisfied,

A. Transient photo excited carrier density should exceed the Coulomb threshold [$n_{crit}^{Coul} \sim 10^{21}\ cm^{-3} \leq n_e$], so the electronic pressure becomes comparable to interlayer cohesive force (36).

B. Induced Coulomb pressure, must exceed the interlayer van der Waals cohesive stress.

C. Pulse duration ($\tau_P$) shorter than e-ph relaxation time, ensuring strong e-lattice non equilibrium [$\tau_P \ll \tau_{e\text{-}ph}$] (38).

D. Deposited energy density remain below the superheating threshold required for phase explosion (39).

## A. Optical field and beam normalization

We first investigate ultrafast carrier excitation in WS$_2$ induced by spatially shaped *fs* laser pulses. The incident electric field is modeled as a linearly polarized pulse of central wavelength 800 nm and temporal duration 50 fs,

$$E(r,t) = E_0 f(r) g(t) e^{-i\omega t} \tag{4}$$

Where f(r) describes the transverse spatial profile under cylindrical symmetry and

$$g(t) = \exp\left(-\left(\frac{t}{\tau_p}\right)^2\right) \tag{5}$$

is the Gaussian temporal envelope. The corresponding intensity distribution is $I(r,t) = I_0 |f(r)|^2 g^2(t)$, with $I_0 = \frac{1}{2} n_0 \varepsilon_0 c |E_0|^2$, where $n_0$ is the refractive index of WS$_2$. Two distinct transverse beam configuration are considered. The Gaussian beam defined as $f_G(r) = \exp\left(-\frac{r^2}{\omega^2}\right)$ (40) and the zeroth order Bessel beam is described by $f_B(r) = J_0(k_r r)$. Where J$_0$ is the Bessel function of the first kind and k$_r$ determines the transverse wave vector of the conical components (41). The Bessel beam redistributes energy into a narrow central lobe surrounded by concentric rings, whereas the Gaussian beam confines energy exponentially near the axis. To ensure the meaningful comparison, both beams are normalized to deliver identical incident fluence,

$$F = \iint_0^\infty I(r,t)\, 2\pi r\, dr\, dt \tag{6}$$

This normalization ensures the isolation of the influence of spatial intensity redistribution on nonlinear carrier excitation.

**B. Nonlinear carrier generation**

For a pulse with longer duration, the absorption process is linear and obeys beer lamberts law. However at higher peak intensities the process is no longer linear (39). So in this case the carrier generation is modelled, to describe how the free e⁻ density evolves under high intense laser irradiation. The simplest way to describe plasma formation is by incorporating three terms, first the field ionization (FI), secondly the Impact ionization and finally the recombination term, which are defined in single rate equation (42).

$$\frac{dn_e}{dt} = W_{SFI}(I(t)) + \alpha I(t) n_e - \frac{n_e}{\tau_r} \qquad (7)$$

Where $W_{SFI}$ describes strong field ionization, $\alpha I n_e$ is for impact (or avalanche) ionization and $\tau_r$ represents carrier relaxation. Since $\tau_p \ll \tau_{e-ph}$, lattice heating during the pulse is negligible and the electronic subsystem stores the majority of the absorbed energy. So, the spatial distribution of intensity $I(r,t)$ is therefore become the critical parameter governing the pathways.

Carrier recombination in semiconductors typically occurs on *ps* timescales, which are much longer than the *50 fs* excitation pulse (43, 44). As a result, carrier losses during the pulse is negligible, and the instantaneous carrier population is primarily governed by generation, so we attribute the rate equation only to generation. Further at high intensities near sub *100 fs* regime, photoionization dominates over impact ionization because it requires the pre-existing free carriers and longer interaction times (42). At a central wavelength of 800 nm, the photon energy $\hbar\omega \approx 1.55\ eV$ is smaller than the bandgap of WS$_2$ ($Eg \approx 2.0\ eV$), thereby supressing single photon interband excitation. The lowest order allowed transition is two photon absorption. Under these conditions, the carrier density evolution is governed by a nonlinear generation rate and the carrier density satisfy,

$$\frac{dn(r,t)}{dt} = \sigma_2 I^2(r,t) \qquad (8)$$

Where $\sigma_2$ denotes the effective two photon absorption co-efficient. Integrating over time yields the residual carrier density after pulse irradiation,

$$n(r) = \sigma_2 \int_{-\infty}^{\infty} I^2(r,t)\, dt \tag{9}$$

The spatial variation in the field profile is amplified by the nonlinear excitation process. Even under identical fluence conditions, difference in peak intensities and spatial localization results in different carrier densities.

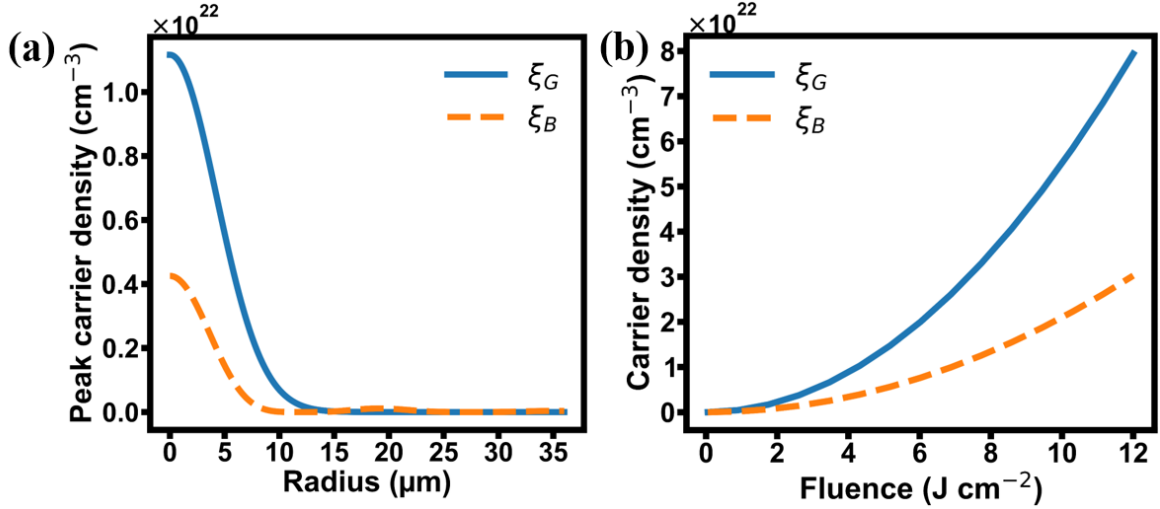

**FIG. 2**: (a) Radial carrier density variation near the peak pulse for both Gaussian and Bessel beams (b) fluence dependent carrier density variation.

The radial carrier density of excited carrier density at the temporal peak of the pulse is shown for both Gaussian ($\xi_G$) and Bessel ($\xi_B$) beam irradiation in Fig. 2(a). The Gaussian pulse exhibits a sharply localized carrier population confined to the focal region, with rapid decay due to its exponential intensity distribution. Even at higher fluence, the excitation remains spatially confined which is leading to steeper carrier density gradients, which can be observed in fluence dependence in Fig. 2(b). The Bessel beam produces an extended carrier distribution. Owing to its non-diffracting central core and concentric ring like structure, higher carrier densities are still sustained over a larger radius. The calculated carrier densities exceed the estimated coulomb instability threshold $n_{crit}^{Coul} \sim 10^{21}\ cm^{-3}$ (45, 46), indicating that electronic destabilization of interlayer bonding is energetically feasible during the pulse irradiation (46). Although both beam profiles generate carrier densities above the Coulomb threshold, the difference in localization leads to difference lattice response. Since the pulse duration ($\tau_P$ = *50 fs*) is much shorter than the electron-phonon coupling time ($\tau_{e\text{-}ph}$), the lattice temperature remains unchanged during the carrier generation, establishing a transient non equilibrium state dominated by electronic excitation. Further the pulse duration ($\tau_p = 50\ fs$) is much shorter than the characteristic electron – phonon coupling time ($\tau_{e-ph}$, typically in few $ps$ in TMDC). This inequality establishes a strong non-equilibrium state in which energy deposition into the electronic system occur faster than energy transfer to the lattice, leading to the build-up of electronic pressure prior to the lattice heating.

## C. Coulomb Stress

To quantify the nonthermal electronic pressure in the ultrafast excitation regime, we estimate the transient carrier density $n_e$ generated under *fs* irradiation for both Bessel and Gaussian beam profiles. The calculated peak densities are $n_e^B \sim 0.41 \times 10^{28}\ m^{-3}$, $n_e^G \sim 1.2 \times 10^{28}\ m^{-3}$ and the associated electronic pressure is approximated as,

$$P_C = \frac{n_e^2\ e^2}{2\varepsilon_0 \varepsilon_r},\ \text{where } \varepsilon_r \approx 15 \text{ is the static dielectric constant of WS}_2.$$

Substituting the above carrier densities onto the electronic pressure yields $P_C^B \approx 0.4 - 0.6$ GPa and $P_C^G \approx 2 - 3$ GPa. The interlayer vdW cohesive stress in layered WS$_2$ is typically on the order of $0.2 - 0.4$ GPa (47). The coulomb pressure induced by the Bessel beam is comparable to the interlayer force in WS$_2$, whereas Gaussian excitation generates electronic stress that significantly exceeds this binding threshold.

## D. Spatial energy density and localization effects

To isolate the role of spatial energy localization independent of total fluence, we evaluate the deposited energy density defined as

$$U(r) = \int_{-\infty}^{\infty} S(r,t)dt \tag{10}$$

Where $S(r,t)$ denotes the spatiotemporal intensity distribution. At an optimized fluence of 3.14 J/cm², the computed peak energy densities are $U_{peak}^{(B)} = 2.02 \times 10^5$ Jcm$^{-2}$ and for $U_{peak}^{(G)} = 5.86 \times 10^6$ Jm$^{-2}$. Thus the Gaussian produces higher ($U_{peak}^{(B)} < U_{peak}^{(G)}$) which arises purely from spatial localization. Because lattice heating is governed by local energy density rather than global fluence, the higher $U_{peak}^{(G)}$ under Gaussian excitation implies a larger transient lattice temperature at the beam center. Consequently, Gaussian irradiation is more likely to drive the material toward a superheated state, whereas the Bessel beam operates in a comparatively distributed excitation regime.

**E. Thermal channel proof**

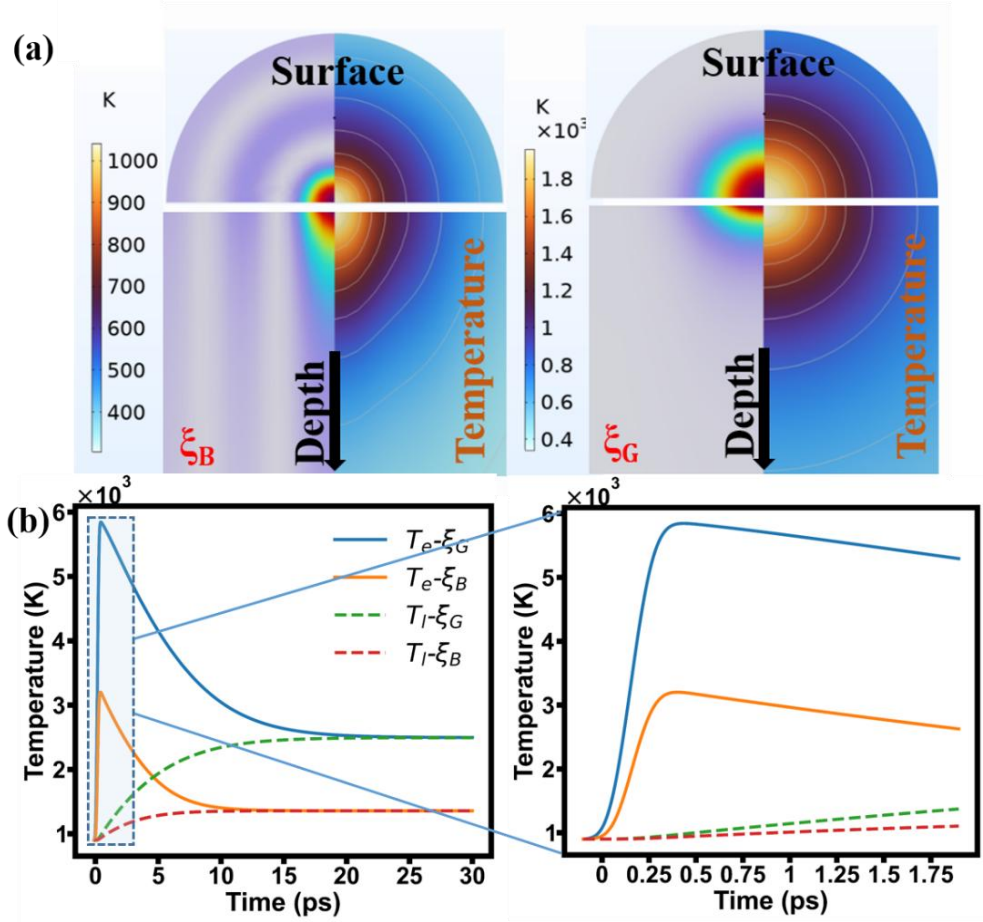

**FIG. 3:** (a) Simulated spatial profile of energy deposition and resulting temperature distribution on the surface and the depth of the material for ξ$_G$ and ξ$_B$ (b) Temporal evolution of electron and lattice temperature following irradiation (Inset highlights the temperature rise within first 2 *ps*)

We analyse the carrier dynamics that are described within the two temperature model (48), where electron and lattice subsystems are characterized by temperatures T$_e$ and T$_l$, respectively. The evolution follows as,

e⁻ System:

$$C_e(T_e) \frac{\partial T_e}{\partial t} = \nabla \cdot (K_e(T_e) \nabla T_e) - G(T_e - T_l) + S(x,t) \quad (11)$$

Lattice System:

$$C_l(T_l) \frac{\partial T_l}{\partial t} = \nabla \cdot (K_l(T_l) \nabla T_l) + G(T_e + T_l) \quad (12)$$

where $C_e$ and $C_l$ are the electron and lattice heat capacities, $K_e$ and $K_l$ are their respective thermal conductivities, G is the electron–phonon coupling constant, and S(x,t) is the laser source term. Under Gaussian excitation the resultant electron temperature reaches to $T_e \approx 5800\ K$, while the lattice assumed initially low as in Fig. 3(b). This extreme non equilibrium ($\Delta T \sim 5500K$) drives rapid electron-phonon energy transfer. The characteristic relaxation time is approximately $\tau_{ep} \approx \frac{C_e}{G}$ (49). As the electron temperature rises to very high value, this further increases electron heat capacity $C_e$. Consequently the electron subsystem stores more energy. Although the coupling term $G(T_e - T_l)$ is large, the increased heat capacity prolongs the relaxation, as more energy to be transferred before reaching the equilibrium. Within few picoseconds, thermal equilibrium is achieved at a temperature of $T_e = T_l \approx 2700\ K$. In case of Bessel beam, considering instantaneous electronic excitation, the electron sub system reaches $T_e = 3300\ K$ while the lattice remains unheated immediately after pulse absorption. Electron-phonon coupling transfers energy to the lattice within several picoseconds reaching an equilibrium temperature of $T_e = T_l \approx 1200\ K$. Although the initial non equilibrium is strong the final lattice temperature is lower than the Gaussian case. Both non-thermal and thermal ablation are inherent under *fs* ablation, however their dominance is beam-profile dependent.

Table 1: Comparative ultrafast excitation parameters revealing beam profile dependent ablation mechanism in $WS_2$

| Parameters | Bessel Beam | Gaussian | Implication |
|---|---|---|---|
| Peak carrier density ($n_e$) | $0.41 \times 10^{28}\ m^{-3}$ | $1.2 \times 10^{28}\ m^{-3}$ | Higher localization |
| Coulomb pressure ($P_C$) | $0.4 - 0.6$ GPa | $2 - 3$ GPa | Comparable interlayer stress |
| Peak electron Temperature ($T_e$) | 3300 K | 5800 K | Strong $T_e$ |
| Peak lattice Temperature ($T_l$) | 1200 K | 2700 K | Exceeds melting threshold of $WS_2$ |
| Energy localization | $2.02\ X\ 10^5$ Jcm$^{-2}$ | $5.86\ X\ 10^6$ Jcm$^{-2}$ | Concentrates energy |
| $\tau_p \ll \tau_{e-ph}$ | Yes | Yes | Establishment of non-equilibrium |
| Dominant mechanism | *coulomb explosion* | *phase explosion* | Ablation pathways |

### III. METHODS

The initial synthesis protocol includes a pellet, prepared by compressing ~600 mg of $WS_2$ powder using a hydraulic palletizer. The compacted pellet was sintered at 120º C for 5 hrs to improve its mechanical stability. The pellet was then immersed in a 3 mL of hexane ($C_6H_{14}$) contained in a 5 mL glass beaker, with liquid level maintained ~5-6 mm above the target surface. The pellet was ablated using a chirped pulse amplification Ti; sapphire oscillator (MICRA, Coherent) producing 50 fs pulses (55-60 nm FWHM spectral bandwidth) at 80MHz and 1 W average power. The amplified output provided an average power of 2.5 W, corresponding to a peak power of approximately 67 GW. The laser beam was focused onto the $WS_2$ pellet using a Plano convex lens (f = 8 cm) to generate a Gaussian beam profile and an axicon lens (apex angle 10º) to generate Bessel beam. Identical fluence conditions were maintained for both beam profile configurations. The focal position was corrected for refractive index mismatch at the air-hexane interface to ensure that the geometric focus coincide with the pellet surface. Fine adjustment along the Z axis was performed using a motorized translation stage. The pellet was oriented normal to the incident laser beam and mounted on a motorized X-Y translation stage (Newport ESP300 controller) to enable uniform ablation. The stage was scanned at speeds of 0.2 and 0.4 mm s$^{-1}$ in a raster pattern to generate periodic lines across the pellet with a 30 min exposure time. Following the ablation, the surrounding hexane solution gradually turned light black, due to the dispersion of exfoliated WS$_2$ nanosheets into the solvent.

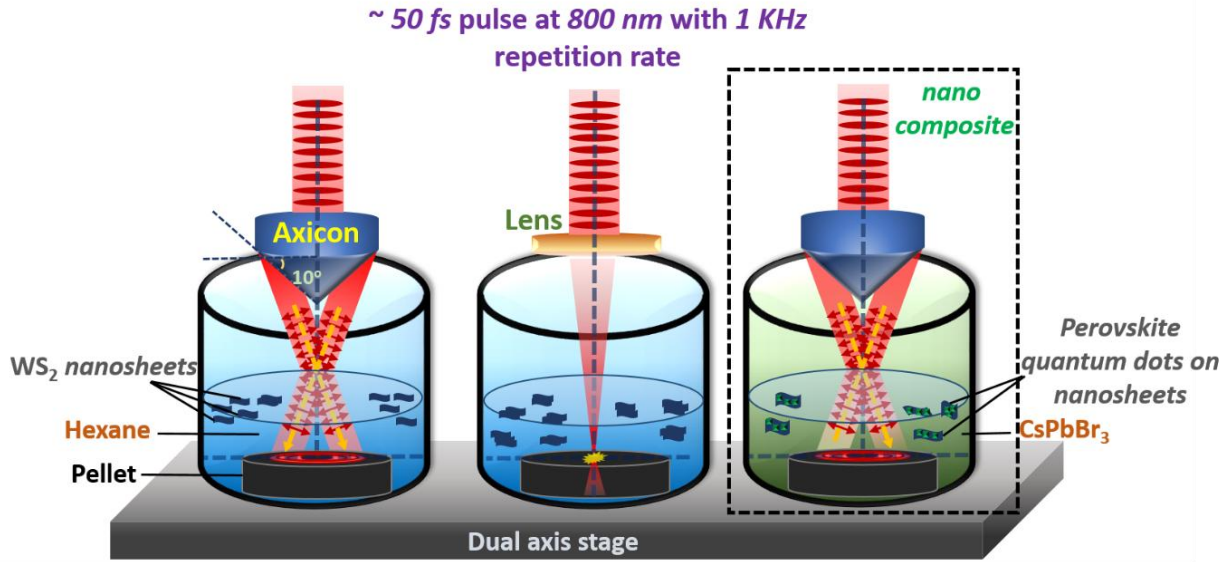

**FIG. 4.** Schematic of laser ablation of WS$_2$ in hexane with Bessel and Gaussian beam profiles, respectively. Highlighted one shows the ablation of WS$_2$ in perovskite dissolved in hexane solution.

## IV. RESULTS AND DISCUSSION

### A. Optical and morphological studies of pristine WS$_2$ ablated with both beam profiles

In the monolayer regime of TMDCs, absence of inversion symmetry and the diminished interlayer coupling, profoundly alters the electronic band structure, phonon dispersion and excitonic behavior relative to their bulk counterparts. Exciting with 488 nm laser is subjugated by 1$^{st}$ order phonon modes, in-plane [$E^1_{2g}(\tau)$] and out of plane $A_{1g}(\tau)$ vibrations, whose frequencies and relative separation evolve with thickness due to change in interlayer restoring forces and dielectric screening. Whereas 532 nm excitation divulge many 2$^{nd}$ order peaks, like double resonance longitudinal acoustic mode (LA) at zone edge activated by disorder showing peak at 352 cm$^{-1}$, merged with in-plane vibration of $E^1_{2g}(\tau)$ mode at 356 cm$^{-1}$ (50). The evolution of $E^1_{2g}(\tau)$ - $A_{1g}(\tau)$ peak separation for synthesized WS$_2$ samples at different conditions is summarized in Fig. 5(a), which decreases substantially after ablation, consistent with reduced restoring force resulting in pronounced redshift of $A_{1g}(\tau)$, while slight blue shift of $E^1_{2g}(\tau)$ observed due to altered dielectric screening (51). Notably, a greater shift in Bessel beam ablated $WS_2$ ($\xi_B$) than Gaussian ablated $WS_2$ ($\xi_G$), implies softening of phonons due to lattice decoupling and reduced anharmonicity corroborating the formation of few layer domains. Ablating material at such high-power laser may result in many heat-affected sheets, to ensure the morphology of $WS_2$ is still constrained, XRD is performed Fig. 5(b). The sample retain the characteristic reflection of 2H-$WS_2$ (space group P63/mmc group), in agreement with JCPDS 08-0237, confirming preservation of hexagonal crystal structure (52). Notably, $\xi_G$ exhibited a greater shift in (002), and broader FWHM is observed in $\xi_B$. Quantitative analysis of dislocation density (calculations are shown in the SI file) yields 4.03 nm$^{-2}$ for $\xi_B$ and 10.95 nm$^{-2}$ for $\xi_G$, demonstrating that $\xi_B$ ablation produces fewer structural defects and a higher density of dislocations or defects induced in $\xi_G$ post ablation (53).

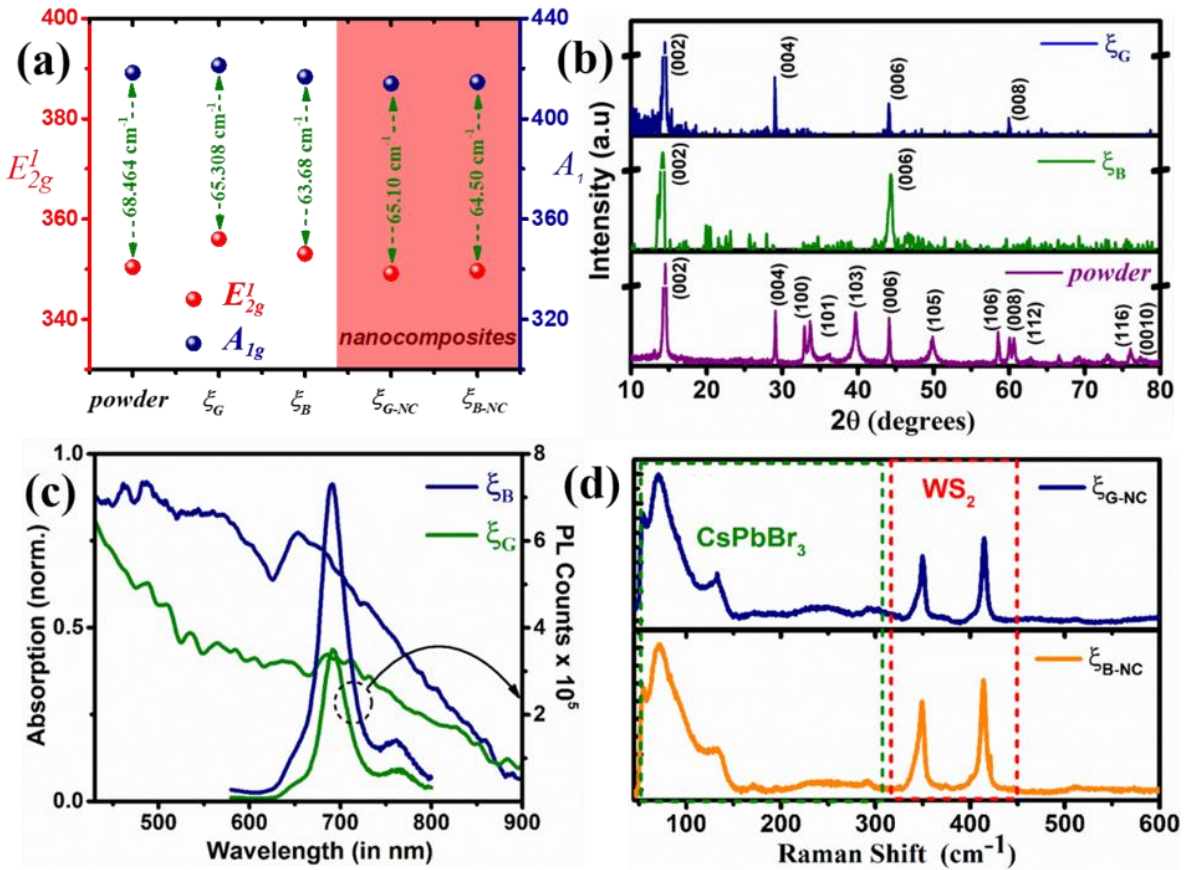

**FIG. 5.** (a) The Raman peak separation of two vibrational modes ($E^1_{2g}$ & $A_{1g}$) in various processed $WS_2$ (b) XRD pattern before and post ablation (c) Normalized absorption and photoluminescence (PL) spectra for $WS_2$ ablated with both beam profiles (d) The characteristic vibrational modes of $CsPbBr_3$ and $WS_2$.

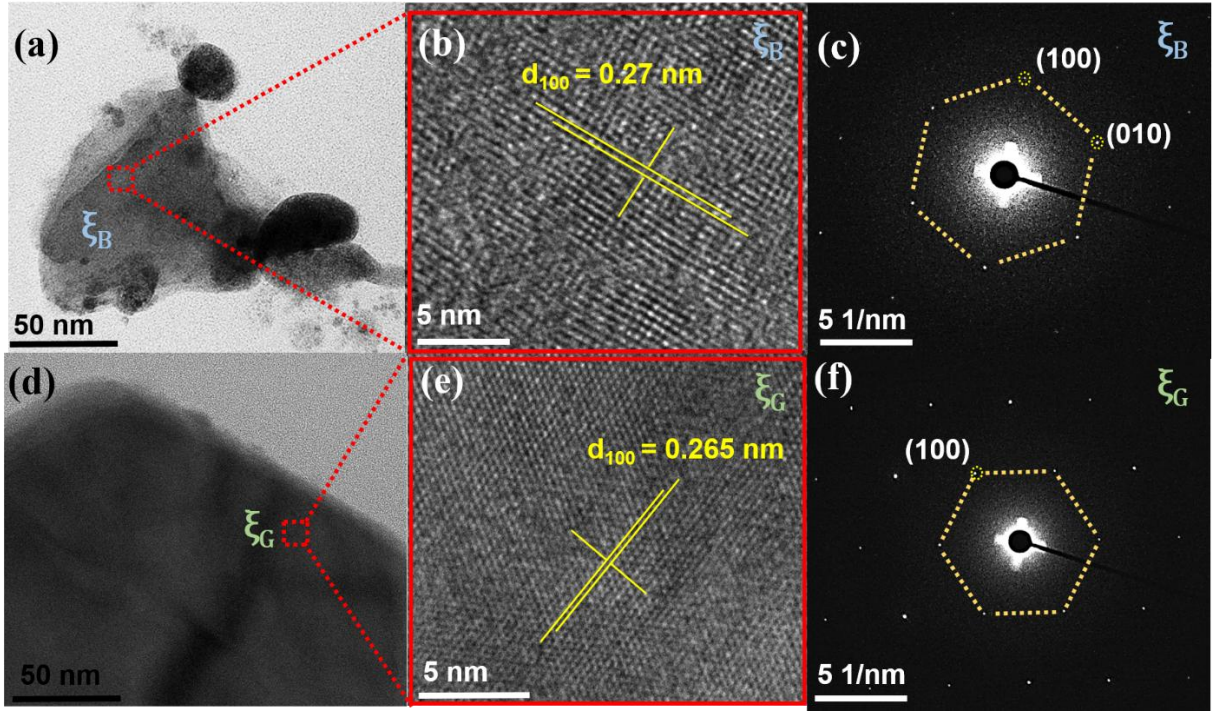

**FIG. 6.** (a), (d) TEM images of ablated WS$_2$ nanosheets with Bessel and Gaussian beam profiles, respectively. (b), (c) Selective Area Electron Diffraction pattern and high-resolution TEM of few layer WS$_2$ with inter planar distance of ~0.27 nm in ξ$_B$ and ξ$_G$ verifying crystallinity intact after laser irradiation. (c), (f) Fourier transform of electron diffraction pattern in ablated WS$_2$ (ξ$_B$ and ξ$_G$)

The absorption has shown similar spectra for both beam profiles in Fig. 5(c), exhibiting three distinct resonances corresponding to the A, B exciton, originating from direct band-edge transition at K (and K′) of the Brillouin zone, split by strong spin-orbit coupling in W d-orbitals which is further enhanced in layered WS$_2$ due to diminished interlayer hybridization (54). The high energy attributed to C exciton arises from interband transition in band nesting region near the Λ-valley, where parallel conduction and valence bands yield a high joint density of states. Fig. 5(c) shows the photoluminescence (PL) spectra normalized with absorption optical density (OD), measured at room temperature under 530 nm excitation. ξ$_B$ revealed enhanced emission attributed to producing a higher proportion of few layer, in contrast ξ$_G$ exhibited lesser intensity consistent with the prevalence of multilayer domains (bi, tri layer, or bulk), where indirect gap recombination competes with direct excitonic emission (55). Spectral deconvolution of PL reveals peaks at 1.79 eV and 1.89 eV associated with direct transition of A, B exciton, additional sub bandgap emission at 1.66 eV is attributed to defect assisted radiative recombination mediated by localized states (56). TEM images presented in Fig. 6(a), 6(d) illustrate the ablated WS$_2$ nanosheets with lateral sizes of 60–80 nm. The SAED pattern demonstrates a honeycomb reciprocal lattice, with diffraction rings indexed to the (100) planes, confirming the 2H phase and in-plane crystallinity.

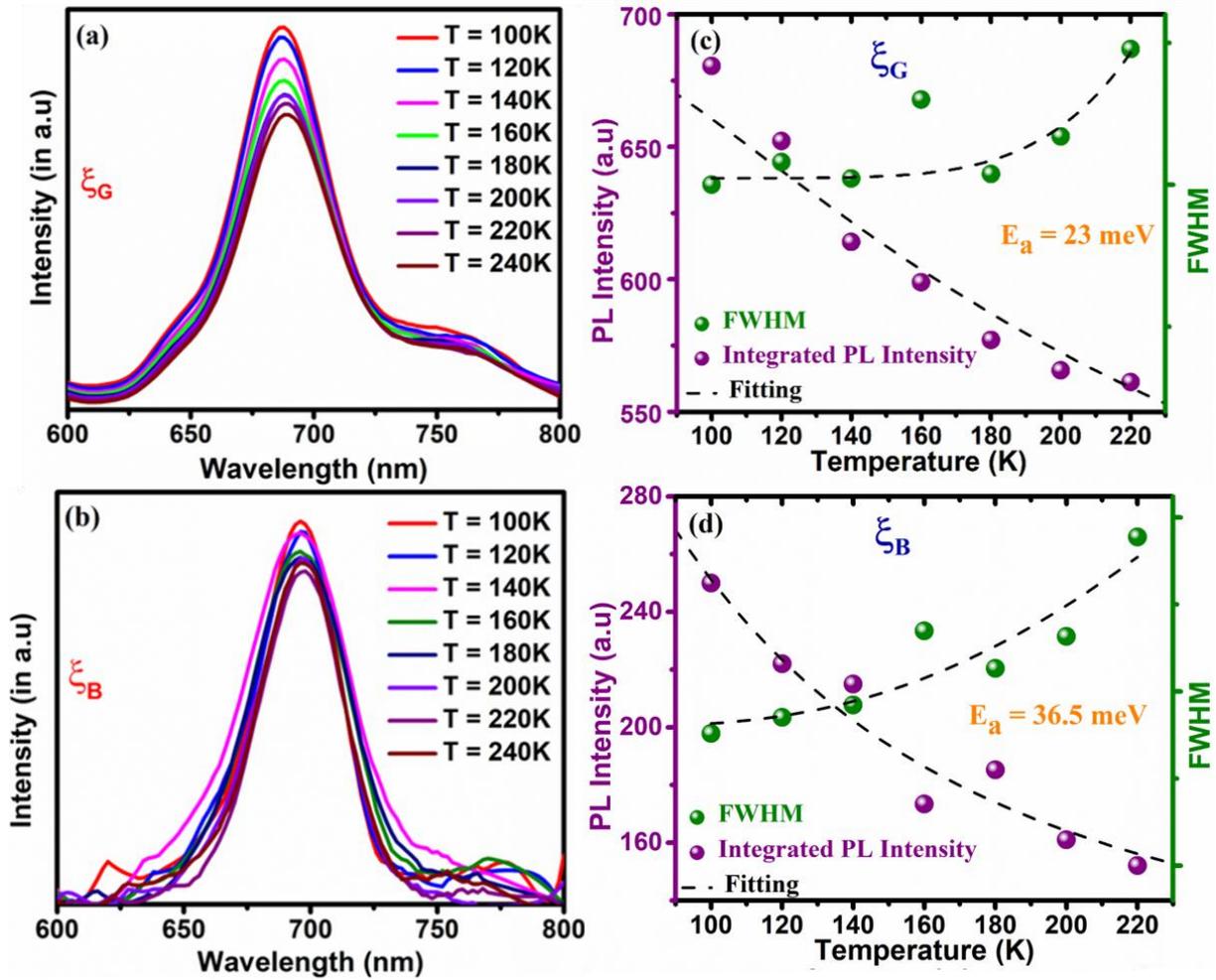

**FIG. 7.** (a), (b) Temperature-dependent PL spectra of $WS_2$ ablated with Gaussian and Bessel beam profiles, respectively. (c), (d) Intensity (left) and FWHM (Right) plots versus temperature. Symbols represent experimental data. Solid lines are fits to the data.

Steady state temperature dependent PL spectra shown in Figs. 7(a), 7(b) have two radiative and a defect assisted recombination peaks at low temperature similar to RT. As of other semiconducting materials, thermally activated non-radiative decay channels with increase in temperature quenches the radiative emission which is observed in Fig. 7. This quenching PL intensity $I(T)$ with varying temperature (T) is well fitted with Arrhenius equation Eq. (13), where $I_0$ is the intensity at 0K , A and $K_b$ are temperature independent pre-exponential and Boltzmann constant, respectively (57).

$$I(T) = \frac{I_0}{1+Ae^{-E_A/K_bT}} \tag{13}$$

The lesser quenching in PL intensity with temperature as in Fig. 7(b) results in higher thermal activation energy of 36.5 meV of $\xi_B$ rendering a stable configuration than 23 meV of $\xi_G$ evaluated by fitting Eq. (13). Further, an increase in the A-exciton luminescence linewidth is observed with temperature which is ascribe to the scattering of optical phonon with exciton. To analyze the environmental interaction of exciton it is fit with Eq. (14) which has contribution of non-phonon ($\Gamma_{np}$) and optical phonon exciton scattering mainly the longitudinal optical phonon ($\Gamma_{LO}$) (58).

$$\Gamma(T) = \Gamma_{np} + \frac{\Gamma_{LO}}{e^{\frac{\hbar\omega}{k_\beta T}}-1} \tag{14}$$

The fitted curve resulted in the estimation of LO phonon energy to be 61.6±10 meV and 30.2±10 meV for $\xi_B$ and $\xi_G$ respectively which is in line with other TMDC materials (59). The higher LO phonon energy extracted for $\xi_B$ reflects the modification of the phonon dispersion arising from reduced interlayer coupling and enhanced in plane lattice stiffness within individual layers (60). The obtained longitudinal optical phonon coupling strength is 609.6±10 meV (for $\xi_B$) and 541±10meV (for $\xi_G$) which is much higher compared to conventional bulk semiconductors, this is due to reduced dimensionality which imparts an additional in-plane confinement to the localised exciton (61). Stronger LO phonon coupling upshot higher probability of phonon assisted non-radiative Auger recombination. The non-radiative broadening term $\Gamma_{np}$ includes multiple excitons scattering channels, including exciton-exciton (62), exciton-carrier, and exciton-defect scattering (63, 64). Since the excitation power is identical in both case, which produces similar exciton density, the contribution of exciton-exciton scattering to $\Gamma_{np}$ remains comparable in both cases. Exciton- carrier is negligible in $WS_2$ due to large exciton binding energy. So we attribute $\Gamma_{np}$ to exciton-defect scattering, these involves surface and substrate defects. (58). A higher $\Gamma_{np}$ of 109.7±10 meV of $\xi_G$ than 94.93±10 meV of $\xi_B$ shows a significant number of defects induced in $WS_2$ while ablating with Gaussian beam. The reduced non radiative broadening in $\xi_B$ indicates that Bessel beam mediated exfoliation suppresses defect assisted exciton scattering while preserving strong exciton-phonon coupling intrinsic to $WS_2$.

## B. Fluence dependent transient absorption spectroscopy of WS₂

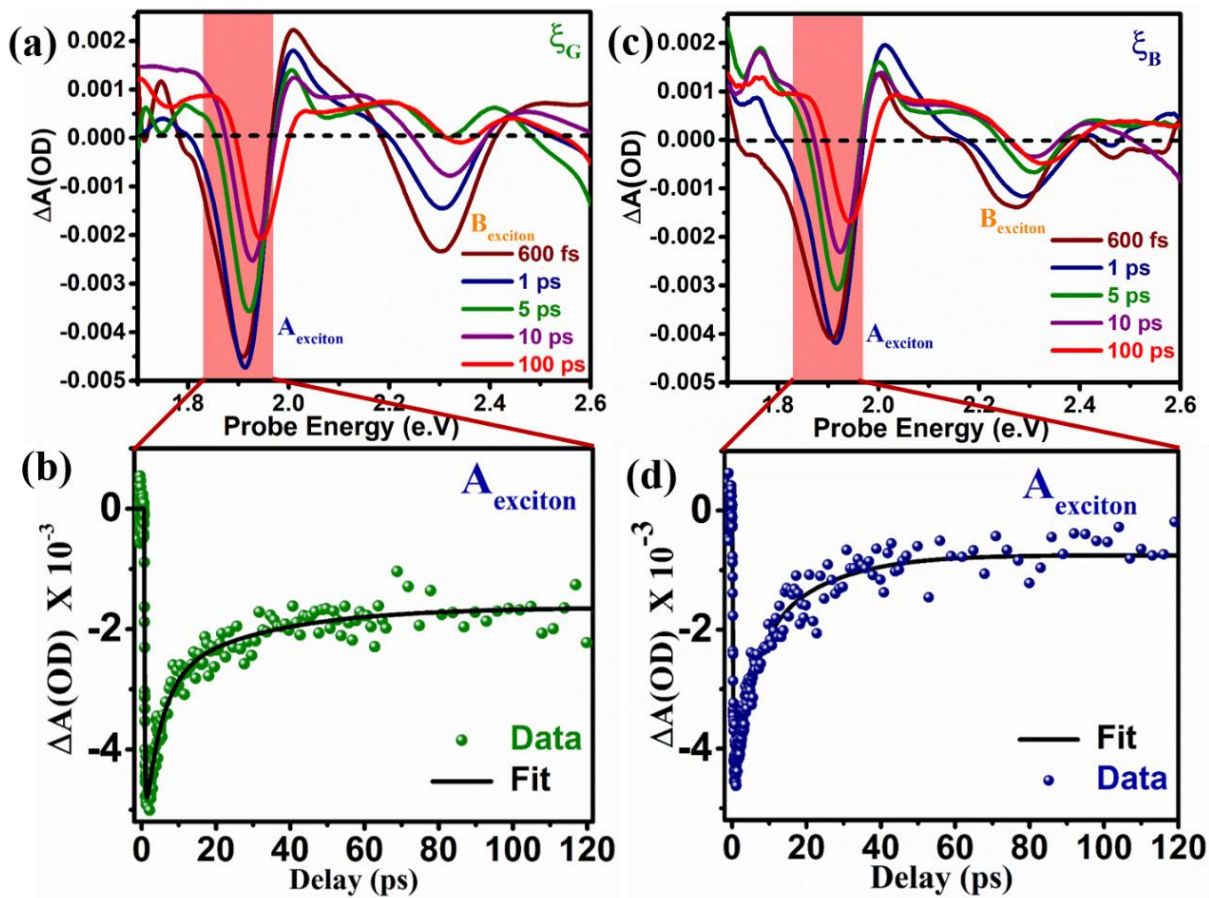

**FIG. 8.** (a), (c) TAS spectra of ablated WS$_2$ at 11.2 µJ/cm$^2$ for ξ$_B$ and ξ$_G$, respectively. (b), (d) the kinetics of A$_{exciton}$ bleach fitted with tri exponential decay function.

The transient absorption spectroscopy (TAS) was employed to elucidate the ultrafast carrier dynamics in layered WS$_2$ under a pump beam of 400 nm (3.1 eV) *fs* excitation. Upon above resonant excitation, electrons are driven from deeper valence bands to higher conduction states. A broad band white light is used to probe the pump induced changes in the molecular system as presented in Fig. 8(a, c), revealing distinct ground state bleach features arising from state filling and Pauli blocking effects at 1.9 eV, 2.3 eV, and 2.6 eV, corresponding to A, B de-convoluted absorption peaks in Fig. 9(a) detailed in the SI. The temporal evolution of the A excitonic bleach is displayed in Fig. 8(b, d) for ξ$_G$ and ξ$_B$ respectively, which is monitored over a range of pump fluence (4-60 µJ/cm²) and was fitted with three exponential decay time constants. The faster component ($\tau_1$) is attributed to surface defect trapping, such as sulphur vacancies and structural dislocations, ($\tau_2$) corresponds to relaxation via emission of optical phonons and ($\tau_3$) captures the band edge carrier recombination, characterizing radiative lifetime of the exciton (65, 66).

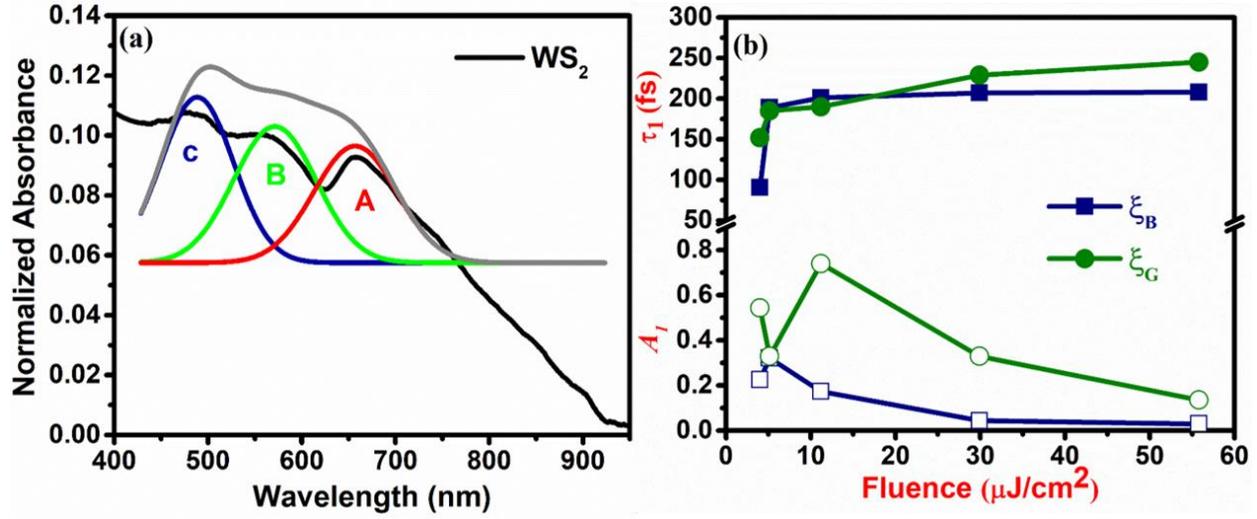

**FIG. 9.** (a) Optical absorption spectra with de-convoluted A, B, C, excitonic peaks of WS$_2$. (c) Evolution of $\tau_1$ (top) and corresponding amplitude (bottom) with increase in pump fluence.

Fluence-dependent analysis of $\tau_1$ in Fig. 9(b) reveals a saturation behavior in both $\xi_B$ and $\xi_G$, indicative of a finite density of trap states. At low pump fluence, a significant fraction of carriers was captured in defect states. However, as the fluence increases, this trap states become progressively filled, diminishing the probability of further trapping and resulting in $\tau_1$ saturation. Notable $\xi_B$ sample exhibits an earlier onset of $\tau_1$ saturation relative to $\xi_G$, suggesting a lower density or shallower energetic distribution of trap states. This behavior points to improved structural coherence or reduced defect formation in $\xi_B$, whereas the delayed saturation in $\xi_G$ indicates a higher density of active trapping centers, likely associated with increased defect-induced disorder. Additionally, the progressive reduction in $\tau_1$ amplitude at higher fluence in Fig. 9(b), reinforces the interpretation that trap-assisted recombination becomes less dominant as available trap states are filled, enabling a larger population of carriers to relax via intrinsic radiative or band-edge recombination pathways.

## C. Ablated WS$_2$ – CsPbBr$_3$ nanocomposites

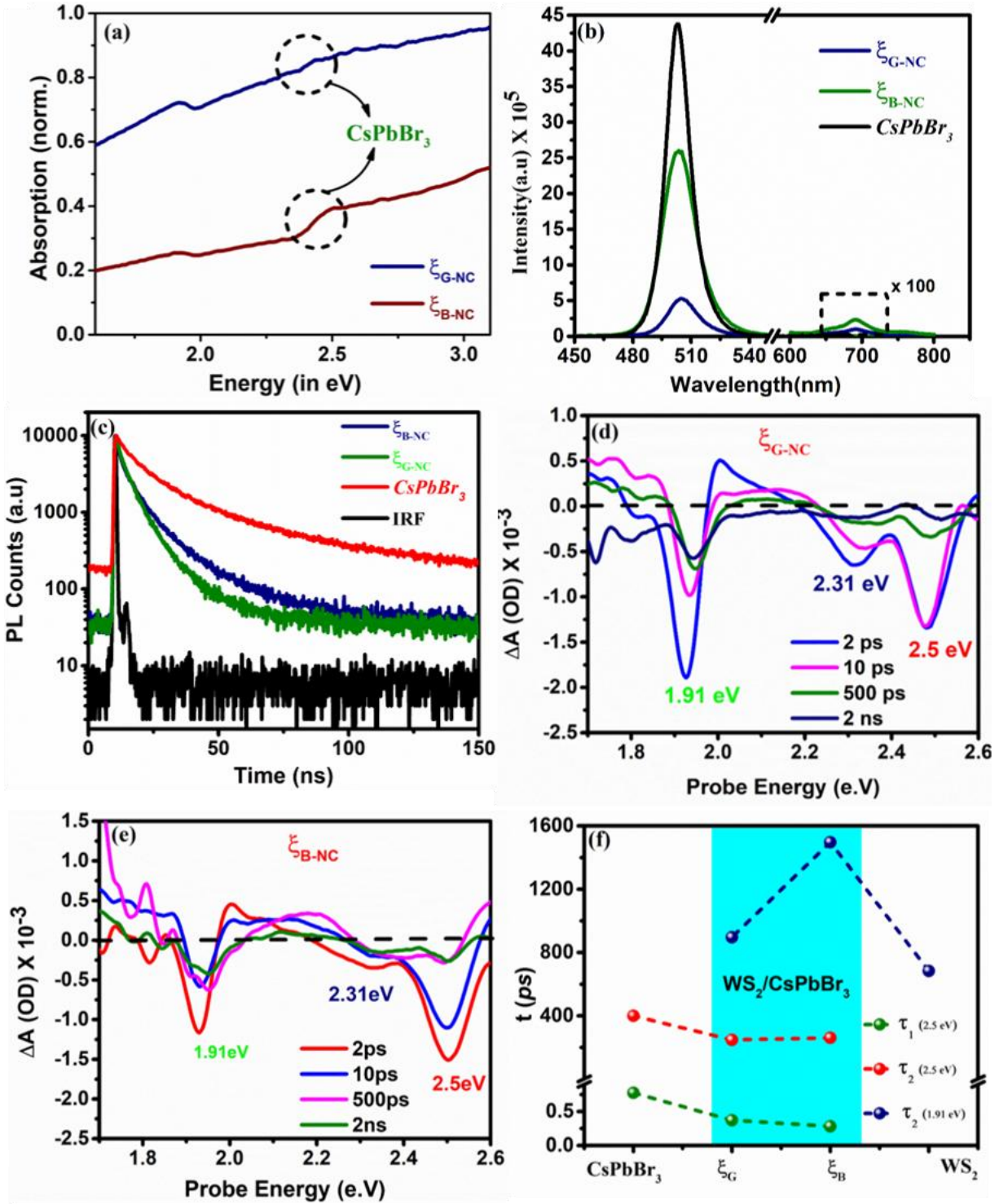

**FIG. 10.** (a) Absorption spectra showing both CsPbBr$_3$ and WS$_2$ contributions (b) Photoluminescence spectra of ablated heterostructure (Inset: focused in TMDC region 600 nm to 800 nm) (c) TrPL monitored at CsPbBr$_3$ emission (d), (e) TAS spectra of NC (f) Variation of $\tau_1, \tau_3$ (of CsPbBr$_3$ bleach) and $\tau_2$ (of WS$_2$ bleach) components in pristine and NCs.

Following the ablation of pristine $WS_2$, we extend this approach in fabricating perovskite nanocomposites. A 2D/0D heterostructure was synthesized with $CsPbBr_3$ quantum dots. The synthesis protocol of quantum dots is followed as in an earlier reported work (67). The $WS_2$ pellet was ablated with perovskites dissolved in hexane solution rather than pure hexane. The laser fluence is optimized such that $WS_2$ is ablated and the perovskite degradation is eschewed. The resulting nanocomposites (NCs) exhibit characteristic absorption edges at ~2.5 eV and ~1.9 eV in the UV–Visible spectra of Fig. 10(a), corresponding to $CsPbBr_3$ and $WS_2$, respectively. PL spectra is noted to perusal the type of hetero-structure NC comprise, Fig. 10(b) shows the subsequent PL quenching of $CsPbBr_3$ with enhancement of $WS_2$ emission, signifying efficient energy transfer from perovskite to TMDC verifying the type I (or straddling) heterojunction at the interface (68). Stronger quenching accompanied by weaker $WS_2$ emission in Gaussian ablated NC ($\xi_{G-NC}$) suggests competing non-radiative pathways which outcompete emission in NC. Time resolved PL Fig. 10(c) further support this interpretation, a tri-exponential fit of emission decay, reveals substantial decrease in carrier lifetime for the NCs: (15.5±0.2) ns for Bessel ablated NC ($\xi_{B-NC}$) and (9.7±0.4) ns for $\xi_{G-NC}$, compared to pristine $CsPbBr_3$ of (18.8±0.3) ns (44). The shortened lifetimes in NCs reflect ultrafast exciton dissociation and charge transfer to $WS_2$, while the stronger lifetime reduction in $\xi_{G-NC}$ implies enhanced non-radiative recombination, likely facilitated by increased structural or surface defects. All fitting parameters are provided in the SI. Additionally, the Raman spectroscopy data presented in Fig. 5(a) reveals a reduced frequency separation in NCs, indicative thinning of $WS_2$ layers, further validating successful exfoliation and nanocomposite integration.

The TAS measurements were performed on the NCs using a low pump fluence of 40 µJ/cm². In contrast to pristine $WS_2$, the NC exhibits an additional bleach feature arising from the contribution of $CsPbBr_3$, which spectrally overlaps with the C-exciton bleach of $WS_2$ as in Fig. 7(d), 7(e). The kinetic fitting of $CsPbBr_3$ bleach at 2.5 eV reveals $\tau_1$, $\tau_2$ components attributing to hot electron and non-radiative relaxation pathways (44). A marked decrease in $\tau_1$ component from $CsPbBr_3$ to NC as in Fig. 10(f) indicates accelerated hot $e^-$ extraction into $WS_2$, which is further pronounced in $\xi_{B-NC}$, demonstrating improved interfacial charge transfer (44). Together the accelerated $\tau_2$ in both NC, implies efficient electron extraction from the perovskite before they recombine radiatively. Similarly, the fitting of A-excitonic bleach of $WS_2$ (at 1.91 eV), reveals $\tau_2$ representing carrier recombination within $WS_2$, which refined in $\xi_{B-NC}$ relative to pristine and $\xi_{G-NC}$, indicating prolonged carrier lifetime, evident by enhanced PL in the inset of Fig. 10(b). Together these findings demonstrates ultrafast energy transfer from perovskite to TMDC, likely facilitated by improved band alignment and interfacial coupling (69). Collectively, the steady state and ultrafast spectroscopic investigation establish a correlation between beam profiles controlled defect engineering and interfacial carrier dynamics in $WS_2/CsPbBr_3$ NC. These observations highlight that the spatial energy distribution during *fs* ablation dictates exfoliation pathways and also governs band alignment, carrier transfer efficiency and heterostructure functionality.

## V. CONCLUSIONS

In summary, we demonstrate that spatial beam profile engineering constitutes a fundamental control parameter in ultrafast laser matter interaction. Through a direct comparison of Gaussian and Bessel beams under identical fluence conditions, we show the redistribution of optical energy in space is sufficient to switch the dominant pathway from thermally driven phase explosion to electronically mediated Coulomb destabilization. This transition originates from distinct interplay between carrier localization, transient electronic pressure and lattice heating induced by beam geometry. Bessel beam excitation generates delocalized and electronically strong interaction volume capable of overcoming interlayer van der Waals cohesion while suppressing excessive lattice heating. The resultant $WS_2$ exhibits reduced defect density, suppressed trap assisted recombination and enhanced excitonic stability. In contrast, Gaussian excitation concentrates energy within a confined focal region, promoting lattice disorder and defect mediated nonradiative decay. Beyond exfoliation, we establish *fs* laser ablation in liquid as a one-step route for the formation of mixed dimensional $WS_2/CsPbBr_3$ nanocomposites. Instead of post synthesis assembly, the heterostructure is obtained in ablation. The defect landscape by beam geometry is shown to directly govern interfacial charge transfer efficiency: Bessel processed systems support efficient carrier extraction, whereas defect rich Gaussian processed structure introduces competing non radiative pathways. By explicitly linking spatial beam geometry to the governing ultrafast energy flow, this work establishes structured light as a decisive control parameter in ultrafast laser matter interaction. This insight moves beyond the conventional pulse parameters-based optimization and offers a generalizable strategy for defect-controlled exfoliation and direct heterostructure formation in layered materials, with direct implications for scalable synthesis of next generation optoelectronic devices.

## Supporting Information

Supporting information includes experimental details, Raman spectroscopy, XRD calculations, TAS of $WS_2$, nanocomposite synthesis and TAS.

## Conflict of Interest

The authors declare no conflict of interest.

## Acknowledgments

CRK acknowledge MOE, IIT Hyderabad. RM acknowledges the financial support from PMRF, India. MSSB acknowledges the ANRF (SERB) India, for NPDF funding (PDF/2023/000958/PMS). V.R. Soma acknowledges the financial support from the DRDO, India through ACRHEM (Phase III) and DIA-CoE. V.R. Soma also acknowledges the support of Director, DIA-CoE, University of Hyderabad.

## Data Availability Statement

The data that support the findings of this study are available from the corresponding author upon reasonable request